\documentclass{PoS}

\title{A global fit to $b\rightarrow c\tau\bar{\nu}$ anomalies after Moriond 2019}

\ShortTitle{A global fit to $b\rightarrow c\tau\bar{\nu}$ anomalies after Moriond 2019}

\author{\speaker{Suman Kumbhakar}\thanks{Thank to IIT Bombay for financial assistant in attending the conference}\\
        Indian Institute of Technology Bombay, Mumbai 400076, India\\
        E-mail: \email{suman@phy.iitb.ac.in}}

\author{Ashutosh Kumar Alok\\
        Indian Institute of Technology Jodhpur, Jodhpur 342037, India\\
        E-mail: \email{akalok@iitj.ac.in}}
        
\author{Dinesh Kumar\\
        National Centre for Nuclear Research, Warsaw, Poland\\
        Department of Physics, University of Rajasthan, Jaipur 302004, India\\
        E-mail: \email{dinesh.kumar@ncbj.gov.pl}}
        
\author{S. Uma Sankar\\
        Indian Institute of Technology Bombay, Mumbai 400076, India\\
        E-mail: \email{uma@phy.iitb.ac.in}}

\abstract{At Moriond 2019, Belle collaboration announced their new measurements on $R_D$ and $R_{D^*}$ which are in agreement with their Standard Model (SM) predictions within $1.2\sigma$. After inclusion of these measurements, the discrepancy between the world averages and the SM predictions of $R_D$-$R_{D^*}$ comes down from $4.1\sigma$ to $3.1\sigma$. Here we do a global fit by taking all measurements in $b\rightarrow c\tau\bar{\nu}$ transition. We find that there are seven allowed new physics solutions, each with different Lorentz structure. Further we show that it is possible to distinguish between them by means of the five observables: the $\tau$ polarization fraction in the decay $B \rightarrow D \tau \bar{\nu}$, the precision measurement of $R_D$, the forward-backward asymmetries in the decays $B \rightarrow (D,D^*) \tau \bar{\nu}$  and the branching ratio of $B_c\rightarrow \tau\bar{\nu}$.}

\FullConference{
European Physical Society Conference on High Energy Physics - EPS-HEP2019 -\\
			10-17 July, 2019\\
			Ghent, Belgium}

\begin{document}

\section{Introduction}

The BaBar, Belle and LHCb collaborations have made a series of measurements of the flavor ratios $R_{D^{(*)}} = \Gamma(B\rightarrow D^{(*)}\,\tau\,\bar{\nu})/ \Gamma(B\rightarrow D^{(*)}\,\{e/\mu\} \, \bar{\nu})$. The average values of these measurements differ from their respective Standard Model (SM) predictions by 4.1$\sigma$, based of HFLAV summer 2018~\cite{Amhis:2016xyh}.  
These measurements indicate the violation of lepton flavor universality. In 2017, LHCb collaboration also measured the related flavor ratio $R_{J/\psi} = \Gamma(B_c\rightarrow J/\psi \, \tau \, \bar{\nu})/\Gamma(B_c\rightarrow J/\psi \, \mu\, \bar{\nu})$ and found it to be $1.7\sigma$ higher than the SM prediction~\cite{Aaij:2017tyk}. In addition to the branching ratios, it is possible to measure various other quantities in $B\rightarrow D^*\,\tau\bar{\nu}$ decay. The polarization fraction of the $\tau$ lepton $P_{\tau}^{D^*}$~\cite{Hirose:2016wfn} has been measured by Belle collaboration which is found to be consistent with the SM prediction. The higher values of $R_D$, $R_{D^*}$ and $R_{J/\psi}$ are assumed to occur due to new physics (NP) contribution to the $b\rightarrow c\,\tau\,\bar{\nu}$ decay. New physics in $b\rightarrow c\, \{e/\mu\}\,\bar{\nu}$ is ruled out by other data~\cite{Alok:2017qsi}.

In the SM, the charged current transition $b\rightarrow c\,\tau\,\bar{\nu}$ occurs at tree level. To account for the measured higher values of flavor ratios, the NP amplitudes are expected to be about $10\%$ of the SM amplitude. The complete list of NP effective operators leading to $b\rightarrow c\,\tau\,\bar{\nu}$ decay are listed in ref.~\cite{Freytsis:2015qca}. These operators can be classified by their Lorentz structure. The purely leptonic decay $B_c\rightarrow \tau\,\bar{\nu}$ is also driven by these operators. This decay mode puts a stronger constraint on the scalar/pseudoscalar NP operators. LEP data imposes a strict upper bound on this quantity which is $\mathcal{B}(B_c\rightarrow \tau\bar{\nu})< 10\%$~\cite{Akeroyd:2017mhr}. In ref.~\cite{Alok:2017qsi}, a global fit was done using the data availble up to summer 2018. The allowed NP solutions, listed in Table IV of ref.~\cite{Alok:2017qsi}, satisfied the constraints (a) $\chi^2_{\rm mim}\leq 4$ and (b) $\mathcal{B}(B_c\rightarrow \tau\bar{\nu})< 10\%$. In ref.~\cite{Alok:2018uft}, a possible discrimination between these NP solutions was studied by means of angular observables in $B\rightarrow (D,D^*)\,\tau\bar{\nu}$ decays. In particular, the tensor NP solution $C_T = 0.516$ can be uniquely distinguished by $D^*$ polarization fraction, if it is measured with a $10\%$ accuracy~\cite{Alok:2016qyh}.

During the last one year, the Belle collaboration has announced two new results: 
\begin{itemize}
\item They made the first measurement of $D^*$ polarization fraction $f_L^{D^*}$. The measured value, $0.60\pm 0.08\,(\rm stat.)\pm 0.04\,(\rm syst.)$~\cite{Adamczyk:2019wyt}, is about $1.5\sigma$  above the SM prediction~\cite{Alok:2016qyh}.
\item At Moriond 2019, they also presented new measurements of $R_D$-$R_{D^*}$, where the $\tau$ lepton was tagged in the decays $B\rightarrow (D,D^*) \,\tau\, \bar{\nu}$. The newly measured values are $R_D = 0.307\pm 0.037\pm 0.016$ and $R_{D^*} = 0.283\pm 0.018\pm 0.014$~\cite{Abdesselam:2019dgh}. These are consistent with the SM predictions~\cite{Amhis:2016xyh}.
\end{itemize}
Including these new measurements in the global averages leads to $R_D = 0.340\pm 0.027\pm 0.013$ and $R_{D^*} = 0.295\pm 0.011\pm 0.008$~\cite{avg19}. The discrepancy between these values and the SM predictions is down to $3.1\sigma$ from $4.1\sigma$. It should be noted that the central values of the new measurements also are higher than the SM predictions, which has been a common feature of all the $R_D$-$R_{D^*}$ measurements.

In this work~\cite{Alok:2019uqc}, we redo the fit to study the impact of these two new measurements on the NP solutions to $b\rightarrow c\tau\bar{\nu}$ anomalies. In section II, we discuss our methodology and fit results. In section III, we describe methods to discriminate between the presently allowed NP solutions. In section IV, we present our conclusions.

\section{New Physics solutions after Moriond 2019}
The most general effective Hamiltonian for $b\rightarrow c\tau\bar{\nu}$ transition, containing all possible Lorentz structures, is \cite{Freytsis:2015qca}
\begin{equation}
H_{\rm eff}= \frac{4 G_F}{\sqrt{2}} V_{cb}\left[\mathcal{O}_{V_L} + \frac{\sqrt{2}}{4 G_F V_{cb}} \frac{1}{\Lambda^2} \left\lbrace \sum_i \left(C_i \mathcal{O}_i +
 C^{'}_i \mathcal{O}^{'}_i + C^{''}_i \mathcal{O}^{''}_i \right) \right\rbrace \right],
\label{effH}
\end{equation}
where $G_F$ is the Fermi coupling constant and $V_{cb}$ is the Cabibbo-Kobayashi-Maskawa (CKM) 
matrix element. Here $\mathcal{O}_{V_L}$ is the SM operator which has the 
usual $(V-A) * (V -A)$ structure. The explicit forms of the four-fermion operators $\mathcal{O}_i$, $\mathcal{O}^{'}_i$ and $\mathcal{O}^{''}_i$ are given in ref.~\cite{Freytsis:2015qca}. We assume that the neutrino is always left chiral. The constants $C_i$ , $C'_i$ and $C''_i$ are
the respective Wilson coeffecients (WCs) of the NP operators in which NP effects are encoded. We set the NP scale $\Lambda$ to be $1$ TeV. Using the effective Hamiltonian
given in eq.~(\ref{effH}), we compute the observables $R_D$, $R_{D^*}$, $R_{J/\psi}$, $P_{\tau}^{D^*}$ and $f_L^{D^*}$ as functions of the various
WCs. By fitting these expressions to the measured values of the observables, we obtain the values of WCs which are consistent with the data. The fit is done by using the CERN minimization code {\tt MINUIT}~\cite{James:1994vla}. The corresponding $\chi^2$ is defined as
\begin{equation}
\chi^2(C_i)= \sum \left(O^{\rm th}(C_i)-O^{\rm exp}\right)^{T} \mathcal{C}^{-1} \left(O^{\rm th}(C_i)-O^{\rm exp}\right).
\label{chi2}
\end{equation}
Here $\mathcal{C}$ is the covariance matrix which includes both theory and experimental correlations.
We perform three types of fits: (a) taking only one NP operator at a time, (b) taking two similar NP operators at a time, (c) taking two dissimilar NP operators at a time. We included the renormalization group (RG) effects in the evolution of the WCs from the scale $\Lambda= 1$ TeV to
the scale $m_b$~\cite{Gonzalez-Alonso:2017iyc}. 
 \begin{table}[htbp]
 \centering
 \tabcolsep 7pt
 \begin{tabular}{|c|c|c|}
\hline\hline
NP type & Best fit value(s)& $\chi^2_{\rm min}$  \\
\hline
SM  & $C_{i}=0$ & $21.80$  \\
\hline
$C_{V_L}$  &  $0.10 \pm 0.02$& $4.5$ \\

\hline
$(C''_{S_L},\, C''_{S_R})$  & $(0.05, 0.24)$ &$4.4$   \\

\hline
$(C_{S_L}, C_T)$ & $(0.06, -0.06)$ & 5.0 \\
\hline
$(C_{S_R}, C_T)$ & $(0.07, -0.05)$ & 4.6 \\
\hline
$(C''_{V_R}, C''_{T})$ & $(0.21, 0.11)$ & 4.2 \\
\hline\hline	

$C_T$ & $-0.07\pm 0.02$ & $7.1$ \\

\hline
$(C'_{V_R},\, C'_{S_L})$  & $(0.38, 0.63)$  &$6.0$  \\
\hline
$(C''_{V_R},\, C''_{S_L})$  & $(0.11, -0.58)$ &$6.2$  \\

\hline\hline
\end{tabular}
\caption{Best fit values of the coefficients of new physics operators at $\Lambda = 1$ TeV by making use of data of $R_D$, $R_{D^*}$, $R_{J/\psi}$, $P_{\tau}^{D^*}$ and $f_L^{D^*}$. In this fit, we use the HFLAV summer 2019 averages of $R_D$-$R_{D^*}$. Here we list the solutions for which $\chi^2_{\rm min}\leq 7$ as well as $\mathcal{B}(B_c\rightarrow \tau\,\bar{\nu})< 10\%$.}
\label{tab1}
 \end{table}

The allowed NP solutions, listed in Table~\ref{tab1}, satisfy the constraints  $\chi^2_{\rm min}\leq 5$ as well as $\mathcal{B}(B_c\rightarrow \tau\bar{\nu})< 10\%$. We note that only the $\mathcal{O}_{V_L}$ solution survives among the single operator solutions. However, its coefficient is reduced by a third. Among the two similar operator solutions, only the $(\mathcal{O}''_{S_L},\, \mathcal{O}''_{S_R})$ persists in principle, with the WCs $(C''_{S_L},\, C''_{S_R}) = (0.05, 0.24)$. The value of $C''_{S_L}$ is quite small, $C''_{S_R} \approx 2C_{V_L}$ and the Fierz transform of $\mathcal{O}''_{S_R}$ is $\mathcal{O}_{V_L}/2$. Therefore, this solution is effectively equivalent to the $\mathcal{O}_{V_L}$ solution. The tensor solution $C_T = 0.516$~\cite{Alok:2017qsi}, which was allowed before $f^{D^*}_L$ measurement, is now completely ruled out at the level of $\sim 5\sigma$.
Table~\ref{tab1} also lists three other solutions with $5\leq \chi^2_{\rm min}\leq 7$. We consider these solutions because the minimum $\chi^2$ is just a little larger than $5$. Hence, they are only mildly disfavoured compare to the five solutions listed above them. One important point to note is that the prediction of $R_D$ (see Table III in ref.~\cite{Alok:2019uqc}) for the tensor NP solution $C_T = -0.07$ is $1.5\sigma$ below the present world average.

\section{Methods to discriminate between new physics solutions}

In order to discriminate between these solutions, we consider the angular observables in $B\rightarrow (D,D^*) \,\tau \,\bar{\nu}$ decays~\cite{Hu:2018veh,Murgui:2019czp,Shi:2019gxi,Blanke:2019qrx}. We consider the following observables:
(i) The $\tau$ polarization $P_{\tau}^D$ in $B\rightarrow D \,\tau \,\bar{\nu}$, (ii) The forward-backward asymmetries $A^{D}_{FB}$ and  $A^{D^*}_{FB}$ in $B\rightarrow (D,D^*) \,\tau\, \bar{\nu}$ and (iii) The branching ratio of $B_c\rightarrow \tau\,\bar{\nu}$. The predictions of each of these quantities for each of the seven solutions are listed in table~\ref{tab2}.
\begin{table}
\centering
\tabcolsep 7pt
\begin{tabular}{|c|c|c|c|c|}
\hline\hline
NP type  & $P_{\tau}^D$ & $A^{D}_{FB}$& $A^{D^*}_{FB}$ &  $\mathcal{B}(B_c\rightarrow \tau\bar{\nu} )\, \%$ \\
\hline
SM & $0.324\pm 0.001$& $0.360\pm 0.001$ & $-0.012\pm 0.007$  & 2.2\\
\hline
$C_{V_L}$  & $0.324\pm 0.002$ & $0.360\pm 0.002$ &$-0.013\pm 0.007$  &2.5\\
\hline
$(C_{S_L}, C_T)$ & $0.442\pm 0.002$ & $0.331\pm 0.003$ & $-0.069\pm 0.009$ & 0.8 \\
\hline
$(C_{S_R}, C_T)$ & $0.450\pm 0.003$ & $0.331\pm 0.002$ & $-0.045\pm 0.007$  &4.0 \\
\hline
$(C''_{V_R}, C''_{T})$ & $0.448\pm 0.002$ & $-0.244\pm 0.003$ & $-0.025\pm 0.008$  & 11.0\\
\hline
\hline
$C_T$ & $0.366\pm 0.003$ & $0.341\pm 0.002$ & $-0.067\pm 0.011$ &  1.9\\

\hline
$(C'_{V_R},\, C'_{S_L})$  &$0.431\pm 0.002$ & $-0.216\pm 0.004$ & $-0.120\pm 0.009$  & 5.7 \\

\hline
$(C''_{V_R},\, C''_{S_L})$  & $0.447\pm 0.003$ & $0.331\pm 0.003$ & $-0.123\pm 0.010$ & 8.4  \\
\hline
\hline
\end{tabular}
\caption{The predictions of $P^D_{\tau}$, $A_{FB}^D$, $A^{D^*}_{FB}$ and $\mathcal{B}(B_c\rightarrow \tau\,\bar{\nu})$ for each of the allowed NP solutions.}
\label{tab2}
\end{table}
From this table we note the following distinguishing features:
\begin{itemize}
\item \underline{$\mathcal{O}_{V_L}$ and $\mathcal{O}_T$ solutions}: The $\mathcal{O}_{V_L}$ and $\mathcal{O}_T$ solutions predict $P_{\tau}^D \approx 0.35$ whereas all the other solutions predict it to be about $0.45$. Therefore a measurement of this observable to a precision of $0.1$ can distinguish these two solutions from the other five. A distinction between the $\mathcal{O}_{V_L}$ and $\mathcal{O}_T$ solutions can be obtained by measuring $R_D$ to a precision of $0.01$, which can be achieved at Belle II~\cite{Kou:2018nap}. 

\item \underline{$(\mathcal{O}''_{V_R}, \mathcal{O}''_{T})$ and $(\mathcal{O}'_{V_R},\, \mathcal{O}'_{S_L})$ solutions}: The $(\mathcal{O}''_{V_R}, \mathcal{O}''_{T})$ and $(\mathcal{O}'_{V_R},\, \mathcal{O}'_{S_L})$ solutions predict $A_{FB}^D$ to be 
$\sim -0.24$ whereas other five solutions predict it to be $\sim 0.33$. Establishing this variable to be negative will distinguish these two solutions from the others. A further distinction between these two solutions can be made  through $A_{FB}^{D^*}$, predicted to be $-0.025$ by 
the $(\mathcal{O}''_{V_R}, \mathcal{O}''_{T})$ solution and $-0.125$ by the $(\mathcal{O}'_{V_R},\, \mathcal{O}'_{S_L})$ solution. 
A measurement of this asymmetry, establishing it to be  either less than or greater than $-0.08$, can distinguish between these two solutions.

\item \underline{The other three solutions}: The three solutions, $(\mathcal{O}_{S_L}, \mathcal{O}_T)$, $(\mathcal{O}_{S_R}, \mathcal{O}_T)$ and $(\mathcal{O}''_{V_R},\, \mathcal{O}''_{S_L})$, all predict the same values for $P_{\tau}^D$ and $A_{FB}^D$. However, the last among them predicts $A^{D^*}_{FB} = -0.123$, whereas the other two predict it to be $\sim -0.05$.  A distinction between the last two solutions and the other two can be made either by establishing this asymmetry to be greater or less than $-0.08$. The respective predictions for $\mathcal{B}(B_c\rightarrow \tau\,\bar{\nu})$ of these three solutions are $0.8\%$, $4.0\%$ and $8.4\%$. Thus a mesaurement of  $\mathcal{B}(B_c\rightarrow \tau\,\bar{\nu})$ to a precision of $2\%$ can distinguish between these three solutions. 
\end{itemize}

\section{Conclusions}
After Moriond 2019, the discrepancy between the the global average values and the SM predictions of $R_D$-$R_{D^*}$ reduces to $3.1~\sigma$. The measured value of $f_L^{D^*}$ rules out the previously allowed tensor NP solution at $\sim 5\sigma$ level. We do a fit with the new global averages and find that there are only {\bf seven} allowed NP solutions. We propose methods to discriminate between these solutions by angular observabes in $B \to (D,D^*) \,\tau\, \bar{\nu}$ decays and the branching
ratio $\mathcal{B}(B_c\rightarrow \tau\,\bar{\nu}$). We find that each of these seven solutions can be uniquely identified by the combination of the five observables with the following described precision:
(i) The $\tau$ polarization $P_{\tau}^D$ in $B\rightarrow D\, \tau \,\bar{\nu}$ to a precision $0.1$, (ii) The ratio $R_D$ to a precision of $0.01$, (iii) The $A_{FB}^D$ in $B\rightarrow D\, \tau\, \bar{\nu}$ to be either positive or negative, (iv) The $A^{D^*}_{FB}$ in $B\rightarrow D^*\,\tau\,\bar{\nu}$ to a precision of $0.02$, and (v) The branching ratio of $B_c\rightarrow \tau\,\bar{\nu}$ to a precision of $2\%$.


\begin{thebibliography}{99}


   \bibitem{Amhis:2016xyh}
  Y.~Amhis {\it et al.} [HFLAV Collaboration],
  Eur.\ Phys.\ J.\ C {\bf 77} (2017) no.12,  895
  
  \bibitem{Aaij:2017tyk}
  R.~Aaij {\it et al.} [LHCb Collaboration],
  Phys.\ Rev.\ Lett.\  {\bf 120} (2018) no.12,  121801
  
   \bibitem{Hirose:2016wfn} 
  S.~Hirose {\it et al.} [Belle Collaboration],
  Phys.\ Rev.\ Lett.\  {\bf 118}, no. 21, 211801 (2017)
  
  \bibitem{Alok:2017qsi}
  A.~K.~Alok, D.~Kumar, J.~Kumar, S.~Kumbhakar and S.~U.~Sankar,
  JHEP {\bf 1809} (2018) 152
  
  
   
    
    \bibitem{Freytsis:2015qca} 
  M.~Freytsis, Z.~Ligeti and J.~T.~Ruderman,
  Phys.\ Rev.\ D {\bf 92}, no. 5, 054018 (2015)
  
  \bibitem{Akeroyd:2017mhr} 
  A.~G.~Akeroyd and C.~H.~Chen,
  Phys.\ Rev.\ D {\bf 96}, no. 7, 075011 (2017)
 
 
  \bibitem{Alok:2018uft} 
  A.~K.~Alok, D.~Kumar, S.~Kumbhakar and S.~Uma Sankar,
  Phys.\ Lett.\ B {\bf 784}, 16 (2018)
  
\bibitem{Alok:2016qyh} 
  A.~K.~Alok, D.~Kumar, S.~Kumbhakar and S.~U.~Sankar,
  Phys.\ Rev.\ D {\bf 95}, no. 11, 115038 (2017)
  
  \bibitem{Adamczyk:2019wyt}
  K.~Adamczyk [Belle and Belle II Collaborations],
  arXiv:1901.06380 [hep-ex].
  
  
  
   \bibitem{Abdesselam:2019dgh} 
  A.~Abdesselam {\it et al.} [Belle Collaboration],
  arXiv:1904.08794 [hep-ex].
  

 
\bibitem{avg19}
  
  https://hflav-eos.web.cern.ch/hflav-eos/semi/spring19/html/RDsDsstar/RDRDs.html
 
\bibitem{Alok:2019uqc}
  A.~K.~Alok, D.~Kumar, S.~Kumbhakar and S.~Uma Sankar,
  arXiv:1903.10486 [hep-ph].
  
\bibitem{James:1994vla}
  F.~James,
  CERN-D-506, CERN-D506.
  
   \bibitem{Gonzalez-Alonso:2017iyc}
  M.~Gonzalez-Alonso, J.~Martin Camalich and K.~Mimouni,
  Phys.\ Lett.\ B {\bf 772} (2017) 777
  
  \bibitem{Hu:2018veh} 
  Q.~Y.~Hu, X.~Q.~Li and Y.~D.~Yang,
  Eur.\ Phys.\ J.\ C {\bf 79}, no. 3, 264 (2019)
  
  \bibitem{Murgui:2019czp} 
  C.~Murgui, A.~Penuelas, M.~Jung and A.~Pich,
  arXiv:1904.09311 [hep-ph].
  
  
  \bibitem{Shi:2019gxi} 
  R.~X.~Shi, L.~S.~Geng, B.~Grinstein, S.~Jager and J.~Martin Camalich,
  arXiv:1905.08498 [hep-ph].
  
  \bibitem{Blanke:2019qrx} 
  M.~Blanke, A.~Crivellin, T.~Kitahara, M.~Moscati, U.~Nierste and I.~Nisandzic,
  arXiv:1905.08253 [hep-ph].
  
  \bibitem{Kou:2018nap} 
  E.~Kou {\it et al.} [Belle-II Collaboration],
  arXiv:1808.10567 [hep-ex].
\end{thebibliography}
\end{document}